\documentclass[aps,prl,twocolumn,groupedaddress,amsmath,nofootinbib]{revtex4-2}
\usepackage{graphicx}
\usepackage{subfigure}
\usepackage{mathtools}
\usepackage{mathrsfs} 
\usepackage{dcolumn}
\usepackage{bm}
\usepackage{amsmath}
\usepackage{color}
\usepackage{verbatim}
\usepackage{natbib} 
\usepackage[dvipsnames]{xcolor}
\usepackage{amsfonts}
\usepackage{soul}
\usepackage{lipsum}
\usepackage{epstopdf}
\usepackage[colorlinks,
            linkcolor=blue,
            anchorcolor=blue,
            citecolor=blue,
urlcolor=blue
            ]{hyperref}
       
\begin{document}

\preprint{APS/123-QED}

\title{\textbf{Approaching the Multiparameter Quantum Cramér-Rao Bound via Classical Correlation and Entangling Measurements} 
}%

\author{Minghao Mi}

\author{Ben Wang}
\email{ben.wang@nju.edu.cn}

\author{Lijian Zhang}
\email{lijian.zhang@nju.edu.cn}

\affiliation{ National Laboratory of Solid State Microstructures, College of Engineering and Applied Sciences, Jiangsu Physical Science Research Center, and Collaborative Innovation Center of Advanced Microstructures, 
Key Laboratory of Intelligent Optical Sensing and Manipulation, Nanjing University, Nanjing 210093, China\\
}


\date{\today}

\begin{abstract}
Multiparameter quantum metrology is essential for a wide range of practical applications. However, simultaneously achieving the ultimate precision for all parameters, as prescribed by the quantum Cramér-Rao bound (QCRB), remains a significant challenge. In this work, we propose a scheme termed local operation with entangling measurements (LOEM) strategy, which leverages classically correlated orthogonal pure states combined with entangling measurements to attain the multiparameter QCRB. We experimentally validate this scheme using a quantum photonic system. Additionally, we employ iterative interactions to demonstrate that the LOEM strategy can achieve the precision of Heisenberg scaling. By theoretically and experimentally demonstrating the saturation of the multiparameter QCRB with the LOEM strategy, our work advances the practical applications of quantum metrology in multiparameter estimation.

\end{abstract}

\maketitle


\textit{Introduction---}Quantum metrology is a cornerstone of quantum science and technology, seeking to enhance measurement precision beyond classical limits by exploiting quantum resources such as entanglement and squeezing~\cite{doi:10.1126/science.1104149,giovannetti_advances_2011,Pirandola2018,https://doi.org/10.1002/qute.202100080,PRXQuantum.3.010202}.~Entanglement can be observed both in the prepared quantum states and in the measurement stage. While single-parameter estimation has been extensively investigated~\cite{Dowling01032008,Slussarenko2017,10.1063/5.0063294}, multiparameter estimation is more relevant for many practical applications, including quantum imaging~\cite{Preza:99,Genovese_2016}, astronomy~\cite{PhysRevA.95.063847,PhysRevA.96.062107}, and sensor networks~\cite{Kómár2014,Nokkala_2018}. Consequently, optimizing the estimation of multiple parameters simultaneously is crucial for maximizing the performance and efficiency of these quantum technologies.

The quantum Cramér-Rao bound (QCRB) establishes the fundamental limit on precision for estimating parameters of a quantum system~\cite{Helstrom1969,holevo2011probabilistic,Xiang2011,doi:10.1126/science.1138007,Yuan2017,Daryanoosh2018}. In single-parameter estimation, performing individual measurements on each probe state, i.e., separable measurements, is sufficient to achieve the QCRB~\cite{PhysRevLett.96.010401}. However, when estimating multiple parameters, the QCRB is not always attainable due to the incompatibility between the optimal measurements corresponding to each parameter~\cite{OZAWA2004367,PhysRevA.94.052108,PhysRevA.105.062442}. This unattainability has motivated the development of various alternative bounds ~\cite{holevo2011probabilistic,PhysRevA.73.052108, 10.1214/13-AOS1147, PhysRevLett.123.200503, Demkowicz-Dobrzański_2020, PhysRevX.11.011028, PhysRevLett.126.120503, PhysRevLett.128.250502,PhysRevResearch.6.033315}, which generally offer lower precisions than the QCRB suggests. Consequently, significant research efforts in recent years have focused on saturating the multiparameter QCRB or the scalar QCRB to achieve higher precision ~\cite{PhysRevLett.127.110501,PhysRevLett.119.130504,PhysRevResearch.5.013138,PhysRevA.97.052127,Chen2024}.

\begin{figure*}[t]
	\centering
	\includegraphics[width=0.99\textwidth]{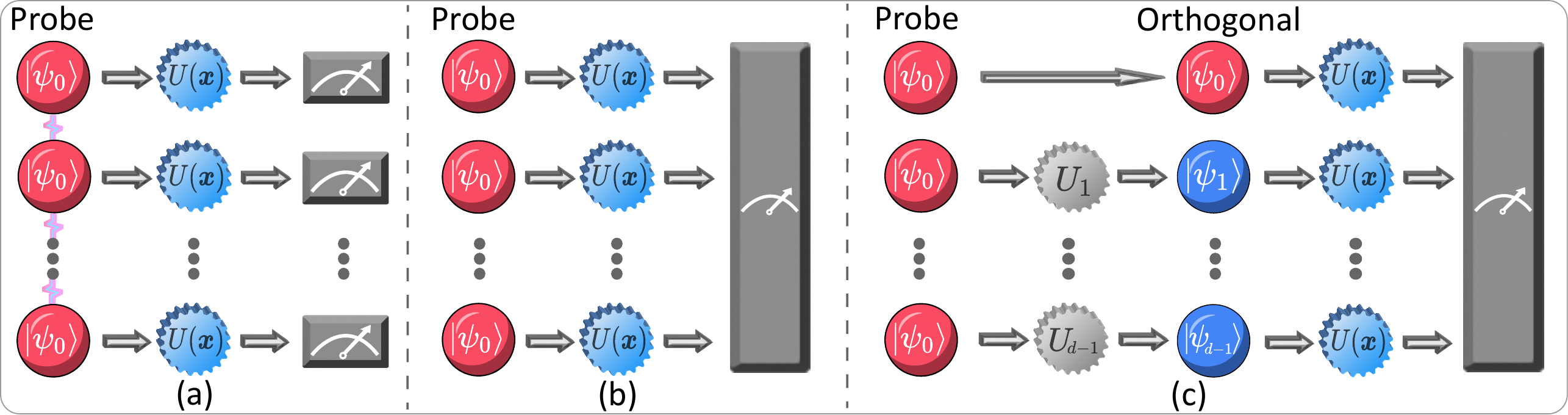}
	\caption{(a) The quantum-classical strategy consists of entangled states and separable measurements. (b) The classical-quantum strategy consists of separable input states and entangling measurements. (c) Our local operation with entangling measurements strategy involves preparing mutually orthogonal states with local operators and performing entangling measurements. }\label{figure1}
\end{figure*}

In single-parameter estimation, the advantage of entanglement is typically confined to the state preparation part [see Fig.~\ref{figure1}(a)]. Entangling measurements performed across multiple copies of probe states [see Fig.~\ref{figure1}(b)], cannot enhance precision in this context~\cite{PhysRevLett.96.010401}. In contrast, multiparameter estimation displays a different behavior, where entangling measurements can indeed improve the precision. However, the impact of entangling measurements varies between pure and mixed states. Recent experimental studies have shown that entangling measurements can enhance parameter estimation precision in mixed states using photonic and superconducting platforms~\cite{Hou2018,Conlon2023}, while these approaches fail to saturate the QCRB. On the other hand, pure states generally provide higher estimation precision than mixed states due to the convexity of the quantum Fisher information matrix (QFIM)~\cite{Liu_2020}. Despite this advantage, entangling measurements on identical copies of parameterized pure states provide no precision enhancement~\cite{KMatsumoto_2002,Conlon2021}. Whether utilizing non-identical, classically correlated pure states for entangling measurements can improve the precision or even achieve the QCRB remains an open question.

In this Letter, we propose a local operation with entangling measurements (LOEM) strategy [see Fig.~\ref{figure1}(c)] that leverages classically correlated states and entangling measurements to saturate the multiparameter QCRB. By encoding the parameters into mutually orthogonal pure states generated through local operations, we achieve the previously unattainable QCRB. {Our work highlights that classical correlations play an important role in multiparameter estimation---a benefit not realized in single-parameter estimation.} To validate this theory, we perform an experimental demonstration on a photonic platform, showing that in qubit systems, the simultaneous estimation of the polar angle $\theta$ and the azimuthal angle $\phi$ can reach the fundamental precision limits. Furthermore, since the variances of $\theta$ and $\phi$ are decoupled under this encoding scheme, iterative interactions can be utilized to approach Heisenberg scaling, which is equivalent to employing a sequential strategy~\cite{Higgins2007,PhysRevLett.128.040503,doi:10.1126/sciadv.adk7616}. Our approach saturates the multiparameter QCRB via classical correlation and entangling measurements, holding significant potential for advancing practical applications of quantum metrology in multiparameter estimation.

\textit{Theoretical framework---}We consider the estimation of $m$ parameters $\boldsymbol{x} = (x_1, x_2, \dots, x_m)^\top$, encoded into the initial pure state $|\psi_0\rangle$ of a $d$-dimensional Hilbert space $\mathcal{H} \cong \mathbb{C}^d$ via the unitary transformation $U(\boldsymbol{x})$, resulting in the parameterized state $|\psi_{\boldsymbol{x}}\rangle = U(\boldsymbol{x}) |\psi_0\rangle$. This parameterized state is subsequently measured using positive operator-valued measures  $\left\{E_k | E_k \geq 0, \sum_k E_k = \mathbb{I}\right\}$. The probability of obtaining outcome $k$ is given by $P(k|\boldsymbol{x}) = \langle
\psi_{\boldsymbol{x}}| E_k|
\psi_{\boldsymbol{x}}\rangle$. From these measurement outcomes, we construct an unbiased estimator 
$\boldsymbol{x}_{\mathrm{est}}$ for the parameters $\boldsymbol{x}$. 
The precision of the multiparameter estimation is quantified by the covariance matrix $\boldsymbol{V}_{\boldsymbol{x}}=\sum_{k}(\boldsymbol{x}_{\mathrm{est}} - \boldsymbol{x})(\boldsymbol{x}_{\mathrm{est}} - \boldsymbol{x})^\top P(k | \boldsymbol{x})$. The Fisher information matrix (FIM), originating from statistical theory~\cite{Fisher_1925}, provides an asymptotic measure of the information content about the parameters. The elements $\mathbf{F}_{i j}$ of the FIM can be calculated as $\mathbf{F}_{i j}=\sum_{k} \partial_{x_{i}}P(k | \boldsymbol{x}) {\partial_{x_{j}}}P(k |\boldsymbol{x})/P(k | \boldsymbol{x})$~\cite{cramer1999mathematical}. The Cramér-Rao bound (CRB) specifies that the covariance matrix of any unbiased estimator is greater than or equal to the inverse of the FIM scaled by the number of experimental runs $M$, i.e.,$\boldsymbol{V}_{\boldsymbol{x}} \geq (\mathbf{F}M)^{-1}$. Although the maximum likelihood estimator (MLE) asymptotically saturates this bound, the CRB is inherently constrained by the chosen measurement strategy. The ultimate precision bound is given by the QCRB, which is based on the quantum Fisher information (QFI) and can be asymptotically saturated in the limit of large $M$~\cite{Helstrom1969,doi:10.1142/S0219749909004839}. The QFI quantifies the maximum amount of information extractable about parameters from a quantum system, optimized over all possible quantum measurements. For a parameterized pure state $|\psi_{\boldsymbol{x}}\rangle$, the QFIM elements are defined as
$\mathbf{Q}_{i j}=4 \operatorname{Re}\left(\left\langle\partial_{x_i} \psi_{\boldsymbol{x}} | \partial_{x_j} \psi_{\boldsymbol{x}}\right\rangle-\left\langle\partial_{x_i} \psi_{\boldsymbol{x}} | \psi_{\boldsymbol{x}}\right\rangle\left\langle\psi_{\boldsymbol{x}} | \partial_{x_j} \psi_{\boldsymbol{x}}\right\rangle\right)$. 
Consequently, the QCRB can be expressed as the following matrix inequality: 
\begin{equation}\label{Bound}
\boldsymbol{V}_{\boldsymbol{x}} \geq (\mathbf{F} M)^{-1} \geq (\mathbf{Q} M)^{-1}.
\end{equation}
Achieving equality in the second inequality requires the vanishing of the mean Uhlmann curvature matrix $\mathcal{U}$, whose elements are given by 
\begin{equation}\label{WCC}
 \mathcal{U}_{ij} = \frac{i}{4} \langle\psi_{\boldsymbol{x}}|[L_i,L_j]|\psi_{\boldsymbol{x}}\rangle,   
\end{equation} 
where $L_{i} = 2(|\partial_{x_i}\psi_{\boldsymbol{x}}\rangle\langle\psi_{\boldsymbol{x}}| + |\psi_{\boldsymbol{x}}\rangle\langle\partial_{x_i}\psi_{\boldsymbol{x}}|)$ is the symmetric logarithmic derivative  operator for the pure state. This condition, known as the weak commutativity condition (WCC)~\cite{10.1063/1.2988130,PhysRevA.94.052108,Vidrighin2014,Carollo2018}, is often difficult to satisfy, generally rendering the QCRB unattainable. In particular, entangling measurements on multiple copies of the pure state $|\psi_{\boldsymbol{x}}\rangle^{\otimes d}$ offer no advantage in multiparameter estimation precision~\cite{KMatsumoto_2002,Conlon2021}.

Instead of employing $d$ identical copies of $|\psi_{\boldsymbol{x}}\rangle$, we consider $d$ distinct states $\{|\psi_{\boldsymbol{x}}^{(j)}\rangle\}_{j=0}^{d-1}$ with classical correlation. Specifically, for a quantum system in a $d$-dimensional Hilbert space $\mathcal{H} \cong \mathbb{C}^d$, the probe state consists of a set of mutually orthogonal pure states $\{|\psi_0\rangle, |\psi_1\rangle, \dots, |\psi_{d-1}\rangle\}$, each of which is a $d$-dimensional state vector in $\mathcal{H}$. These states are generated from the initial state $|\psi_0\rangle$ by applying local unitary operators $U_i$ such that $
|\psi_i\rangle = U_i |\psi_0\rangle$
[see Fig.~\ref{figure1}(c)]. We encode $m$ parameters by applying the same unitary $U(\boldsymbol{x})$ to each state, {with $m \leq 2d-2$ since a single pure qudit state can encode at most $2d-2$ independent parameters.} This process results in the parameterized state $ |\Psi_{\boldsymbol{x}}\rangle = U(\boldsymbol{x})^{\otimes d} \left( |\psi_0\rangle \otimes |\psi_1\rangle \otimes \dots \otimes |\psi_{d-1}\rangle \right) $, where the system is described in a ${d^d}$-dimensional Hilbert space $\mathcal{H}^{ \otimes d}$. This encoding scheme satisfies the WCC, as evidenced by $\text{Im} (\langle\partial_{x_i}\Psi_{\boldsymbol{x}} | \partial_{x_j}\Psi_{\boldsymbol{x}}\rangle) = 0$ for any $i,j$, which is consistent with Eq.~(\ref{WCC}). Further details are provided in Supplementary Material~\cite{LOEM2025_supplementary}. Importantly, the state $ |\Psi_{\boldsymbol{x}}\rangle $ exhibits a global antiunitary symmetry~\cite{miyazaki2020symmetry}. This symmetry allows us to express the quantum state as a superposition of basis states with real coefficients, i.e., $|\Psi_{\boldsymbol{x}}\rangle = \sum_k a_k |k\rangle$, where $|k\rangle$ represents a basis state in the $ {d^d}$-dimensional Hilbert space $\mathcal{H}^{ \otimes d}$, and $a_k$ are real coefficients.
As a result, it becomes straightforward to construct a set of optimal measurements defined by $E_k = |k\rangle\langle k| $, which generally require entangling measurements~\cite{PhysRevLett.133.210801}. Consequently, our LOEM strategy saturates the QCRB for the pure state $ |\Psi_{\boldsymbol{x}}\rangle$.  {Moreover, an additional advantage of our strategy is that it permits the encoding of more than $2d-2$ independent parameters. In particular, since  $\{|\psi_i\rangle\}_{i=0}^{d-1}$ forms a complete basis for the $d$-dimensional Hilbert space, it enables the estimation of up to $d^2-d$ parameters~\cite{LOEM2025_supplementary}.}

\begin{figure}
	\centering
	\includegraphics[width=0.48\textwidth]{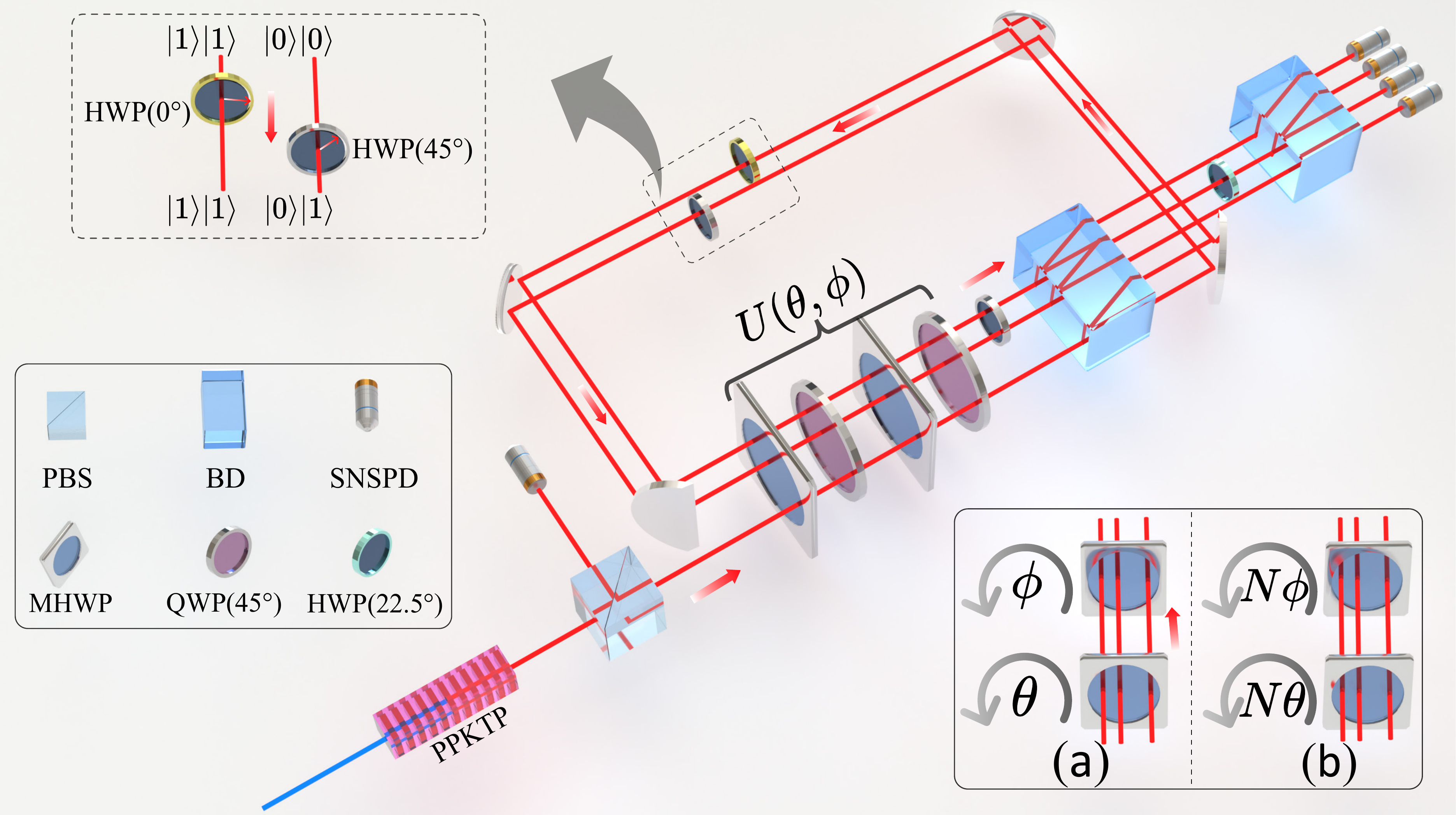}
	\caption{The experimental setup for the joint estimation of $\theta$ and $\phi$ using the LOEM strategy. The setup involves the preparation of mutually orthogonal states that encode the parameters within a loop, followed by an optimal entangling measurement procedure. The inset (dashed box) details the process of making the second qubit orthogonal to the first qubit. Inset (a) depicts the experimental setup used to validate the saturation of the QCRB for $\theta$ and $\phi$, while inset (b) demonstrates the experimental setup for verifying the achievement of Heisenberg scaling.}\label{figure2}
\end{figure}

As an example, we consider two-parameter estimation in qubit systems. A pure state $|\mathbf{n}\rangle$ can be parameterized using spherical coordinates, with the polar angle $\theta \in [0,\pi)$ and the azimuthal angle $\phi \in [0,2\pi)$. The state is given by $|\mathbf{n}\rangle = \cos(\theta/2)|0\rangle + e^{i\phi}\sin(\theta/2)|1\rangle$, which can be obtained by applying the unitary transformation $U(\theta,\phi)$ to the $|0\rangle$ state: $|\mathbf{n}\rangle = U(\theta,\phi)|0\rangle.~ ${Here, the unitary transformation is defined as $U(\theta, \phi) = (\cos {\theta}/{2}, -e^{-i\phi} \sin {\theta}/{2}; e^{i\phi} \sin {\theta}/{2}, \cos {\theta}/{2})$~\cite{LOEM2025_supplementary}}. Neither the single-copy state $|\mathbf{n}\rangle$ nor the two-copy state $|\mathbf{n},\mathbf{n}\rangle$ satisfies the WCC for the parameters $\theta$ and $\phi$. However, by encoding the parameters into mutually orthogonal states, specifically $U(\theta,\phi)|0\rangle \otimes U(\theta,\phi)|1\rangle$, the WCC is satisfied. This strategy produces the state $|\mathbf{n},-\mathbf{n}\rangle$, also known as the antiparallel state, which has been successfully applied to enhance state discrimination~\cite{PhysRevA.59.1070,PhysRevLett.94.220406,PhysRevLett.83.432,PhysRevA.89.042110,PhysRevLett.124.060502}. Notably, since the QFI is the same for $|\mathbf{n}\rangle$ and $|-\mathbf{n}\rangle$,  the QFIM for $|\mathbf{n},-\mathbf{n}\rangle$ is given by

\begin{equation}\label{QFIm1}
    \mathbf{Q}(|\mathbf{n},-\mathbf{n}\rangle)=\left(\begin{array}{cc}2 & 0 \\ 0 & 2 \sin ^{2} \theta\end{array}\right)= 2 \cdot \mathbf{Q}(|\mathbf{n}\rangle).
\end{equation}

\begin{figure}[t]
	\centering
	\includegraphics[width=\linewidth]{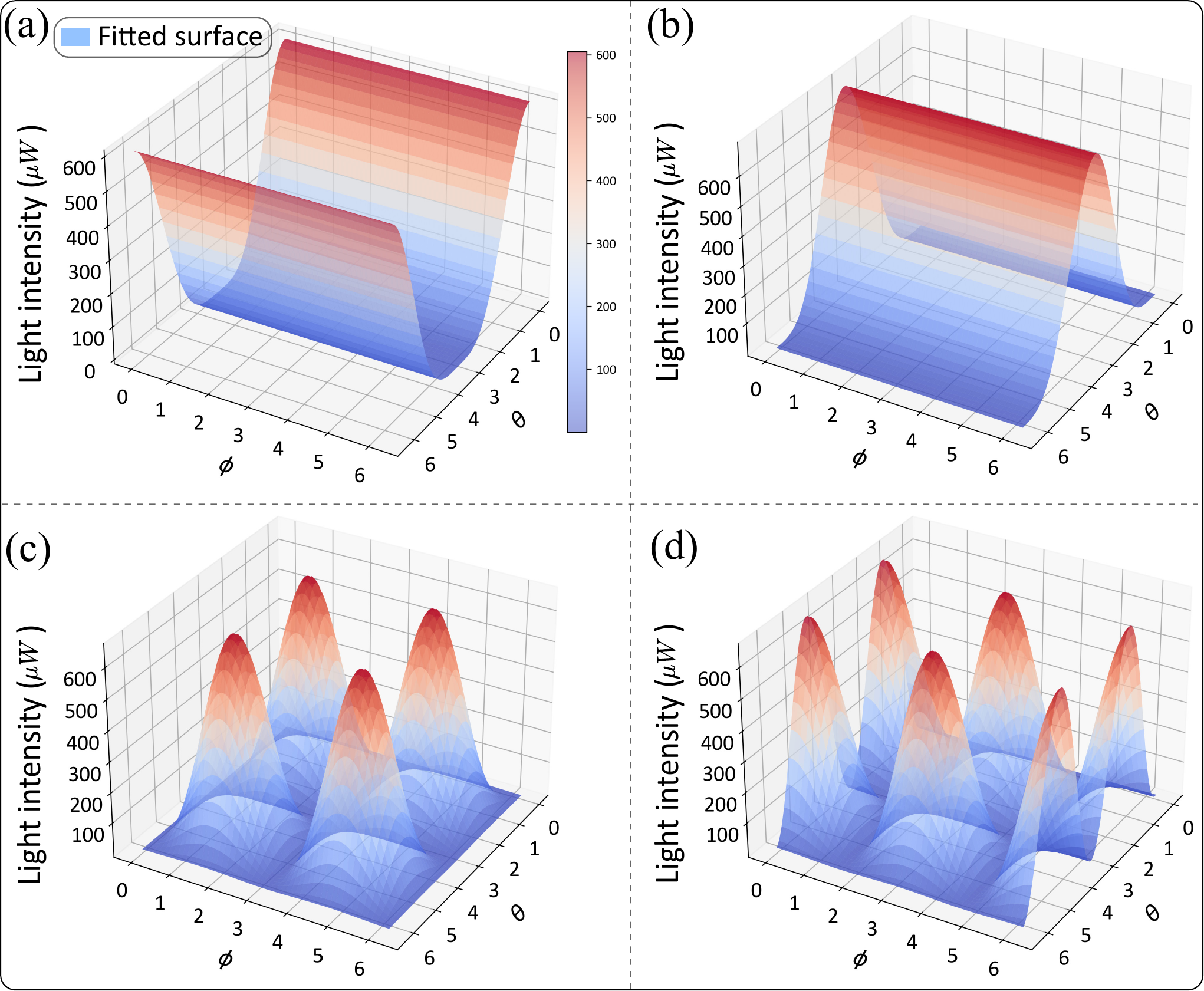}
	\caption{Experimental results of the light intensity distribution across the four output ports as a function of both $\theta$ and $\phi$, where $\theta, \phi \in [0, 2\pi)$. The surfaces in each panel are constructed from a fit to a $100 \times 100$ grid of sampled data points. Panels (a)–(d) correspond to the four output ports, associated with the likelihood functions $P_{1}$, $P_{2}$, $P_{3}$, and $P_{4}$, respectively, evaluated at $N=1$. The experimental results agree well with the theoretical predictions.}\label{likelihood}
\end{figure}

In multiparameter estimation, the FIM satisfies the inequality $(\boldsymbol{F}^{-1})_{ii} \geq (\boldsymbol{Q}^{-1})_{ii} \geq 1/\boldsymbol{Q}_{ii}$. Equality in the first inequality holds if and only if the parameterized quantum states satisfy the WCC. Furthermore, equality in the second inequality is achieved when the quantum QFIM $\boldsymbol{Q}$ is diagonal, indicating that the precision of different parameters does not interfere with one another. In other words, the variances of different parameters are decoupled. Our LOEM strategy simultaneously satisfies the WCC and the diagonal condition. As a result, the variances of the parameters $ \theta $ and $ \phi $ can approach the QCRB and are decoupled. {This allows for an iterative interaction strategy, where the unitary transformation is defined as $U(N\theta, N\phi)$~\cite{LOEM2025_supplementary}, which can be physically realized through adaptive strategy~\cite{Higgins2007,RevModPhys.90.035006,PhysRevLett.123.040501,doi:10.1126/sciadv.abd2986}.} The resulting quantum state is $U (N \theta, N \phi)|0\rangle \otimes U (N \theta, N \phi)|1\rangle$, and the corresponding QFIM is given by
\begin{equation}\label{QFIm2}
    \mathbf{Q}'=\left(\begin{array}{cc}2N^2& 0 \\ 0 & 2N^2 \sin ^{2} (N\theta)\end{array}\right).
\end{equation}
This implies that the variances of $\theta$ and $\phi$ exhibit Heisenberg scaling of $1/N^2$. The optimal measurement for this state is realized by projective measurements onto the basis composed of the four states $|01\rangle$, $|10\rangle$, $(|00\rangle + |11\rangle)/\sqrt{2}$, and $(|00\rangle - |11\rangle)/\sqrt{2}$, which constitute a set of entangling measurements independent of $N$. The measurement outcome probabilities for the four output ports are given by $ P_1=\cos^{4} (N\theta/2) , P_2=\sin^{4}(N\theta/2) , P_3=\sin^{2} (N\theta) \sin^{2} (N\phi)/2 , P_4= \sin^{2} (N\theta)\cos^{2} (N\phi) /2.$  {Furthermore, it can be calculated that under this measurement scheme, the FIM equals the QFIM, thereby saturating the QCRB~\cite{LOEM2025_supplementary}.}

\textit{Experimental setup and results---}The experimental setup for jointly estimating the parameters $\theta$ and $\phi$, encoded in mutually orthogonal states, is shown in Fig.\ref{figure2}. The process consists of preparing the parameterized mutually orthogonal states, followed by performing optimal entangling measurements.

A pair of 1560 nm photons is generated via type-II spontaneous parametric down-conversion (SPDC) in a periodically poled potassium titanyl phosphate (PPKTP) crystal and subsequently separated into two distinct paths using a polarizing beam splitter (PBS). The vertically polarized ($V$) photon is detected by a superconducting nanowire single-photon detector (SNSPD) and serves as a herald. Meanwhile, the horizontally polarized ($H$) photon is used to prepare the parameterized mutually orthogonal states and perform entangling measurements.

\begin{figure}[t]
	\centering
	\includegraphics[width=\linewidth]{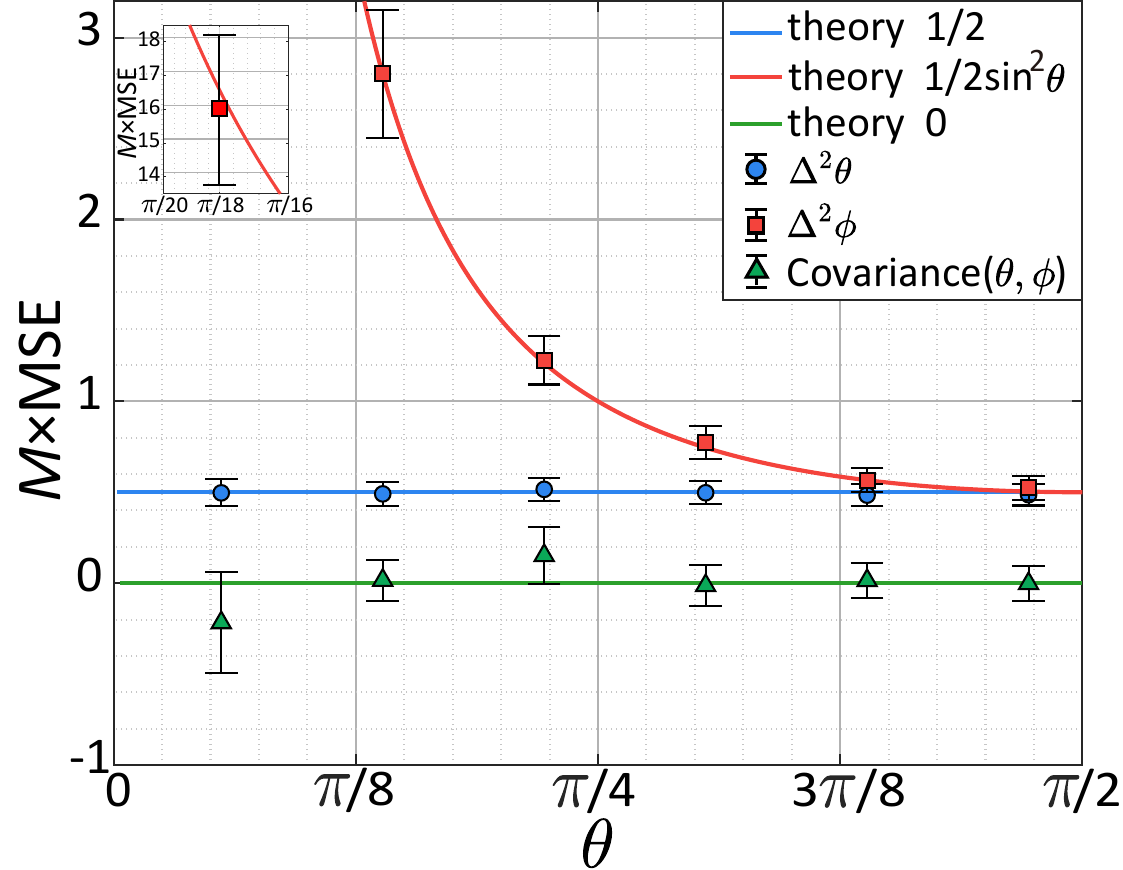}
	\caption{Experimental results demonstrate saturation of the QCRB for the estimation of $\theta$ and $\phi$. The vertical axis represents the product of the number of experimental runs and the mean squared error ($M \times \text{MSE}$),  { where $M$ is approximately $10^4$}. The covariance between $\theta$ and $\phi$ is zero, indicating that the variances of $\theta$ and $\phi$ are decoupled.}\label{result1}
\end{figure}

The first qubit is encoded in the photon's path degree of freedom (DOF), where $|\text{up}\rangle \equiv |0\rangle $ and $|\text{down}\rangle \equiv |1 \rangle $. The second qubit is encoded in the polarization DOF, with $|H\rangle \equiv |0\rangle $ and $|V \rangle \equiv |1 \rangle $. The horizontally polarized photon $|H\rangle$ undergoes a unitary transformation $U(\theta, \phi)$ implemented with a series of wave plates to encode the unknown parameters $\theta$ and $\phi$. Specifically, a rotatable motorized half-wave plate (MHWP) set to a rotation angle of $\theta/4$ is used to encode the parameter $\theta$. Following this, a phase shifter—consisting of a rotatable MHWP at an angle of $(\phi+\pi)/4$ placed between two quarter-wave plates (QWPs) fixed at $\pi/4$—is employed to encode the parameter $\phi$. Subsequently, a beam displacer (BD) shifts the horizontal polarization $|H\rangle$ into the up path $|\text{up}\rangle$, creating a 4 mm separation from the vertical polarization $|V\rangle$ in the down path $|\text{down}\rangle$. This process transforms the polarization qubit into the path qubit. Two HWPs, set to $\pi/4$ and $0$, are then placed in the $|\text{up}\rangle$ and $|\text{down}\rangle$ paths, respectively, to prepare the photon in the state $|\mathbf{n},V\rangle$. To encode the second qubit, a loop is constructed using mirrors such that the photon can undergo the identical unitary transformation as the first qubit, ultimately preparing $|\mathbf{n},-\mathbf{n}\rangle$. The antiparallel state is then directed to the entangling measurements module, which performs optimal measurements based on a two-step quantum walk.

\begin{figure}[t]
        \vspace{0mm}
	\centering
	\includegraphics[width=1\linewidth]{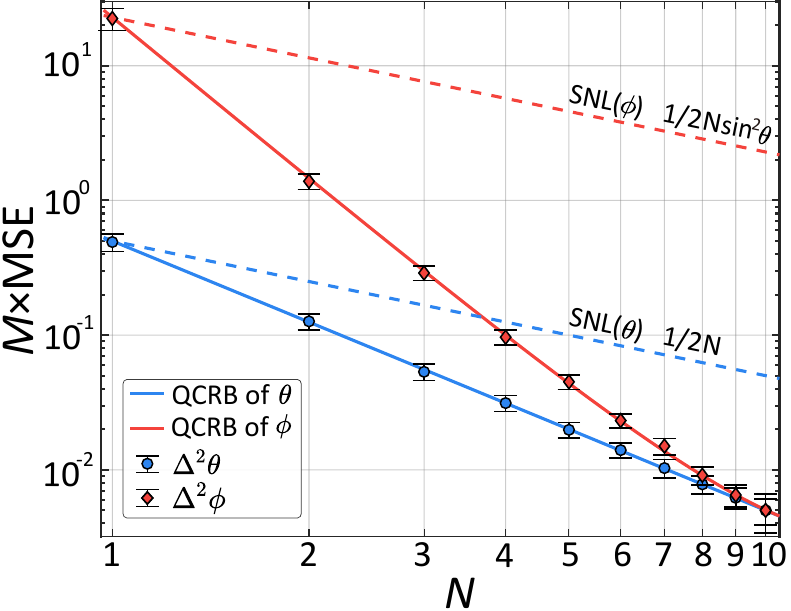}
	\caption{Experimental results demonstrate Heisenberg scaling in $M \times \text{MSE}$  {(with $M \approx 10^4$)} for the simultaneous estimation of ($\theta$, $\phi$). Both $\phi$ and $\theta$ are fixed at 8.5°. $N\theta$ and $N\phi$ are encoded using $N$ iterative interactions. The resulting product $M \times \text{MSE}$ is plotted against the iteration number $N$ on a log-log scale, in comparison with the corresponding shot-noise limits (SNL).}\label{result2}
\end{figure}

The experimental setup is initially calibrated using a 1560 nm classical light source. Light intensity distributions are measured at four output ports as functions of $\theta$ and $\phi$ over the range of 0 to $2\pi$. These distributions closely match the theoretical measurement outcome probabilities (see Fig.~\ref{likelihood}), verifying the accuracy of the setup. The system is then switched to a single-photon source, and photon  counts are recorded at the four output ports using SNSPDs.  {Based on the observed photon--count distributions, we employ a maximum likelihood estimator to simultaneously estimate both $\theta$ and $\phi$~\cite{LOEM2025_supplementary}.}

We first experimentally demonstrate, for the case of $N = 1$, that the LOEM strategy saturates the multiparameter QCRB, as illustrated in the inset (a) of Fig.~\ref{figure2}. Because the QFIM in Eq.~(\ref{QFIm1}) is independent of the parameter $\phi$, we fix $\phi$ at 36° and vary $\theta$ across a set of values: 10°, 25°, 40°, 55°, 70°, and 85°. These define six parameter estimation sets. For each set of parameters, 400 repeated measurements are performed, yielding 400 joint estimates of $\theta$ and $\phi$. From these joint estimates, the estimation variances of $\theta$ and $\phi$ are calculated and compared with the inverse of the QFIM. The same method is then employed to conduct experimental measurements for the other five parameter sets. The experimental results depicted in Fig.~\ref{result1} show that the product of the number of experimental runs and the mean squared error ($M \times \text{MSE}$) closely matches the theoretical predictions.  {Error bars are determined via Monte Carlo simulations using 100 independent samples drawn from the corresponding Poisson distributions~\cite{Carolan2014,Vidrighin2014,LOEM2025_supplementary}.} Moreover, the experiments reveal zero covariance between $\theta$ and $\phi$, confirming that the precision for one parameter does not affect the other.

{We then experimentally demonstrate that the iterative interaction strategy can approach Heisenberg scaling. In particular, the effect of 
$N$ iterative interactions results in an $N$-fold amplification of the parameters $\theta$ and $\phi$, corresponding to the encoded unitary transformation
$ U(N\theta, N\phi) = U(N\phi) \, U(N\theta)$ with the constraints $0 \leq \phi < {\pi}/{2N}$ and $0 \leq \theta < {\pi}/{2N}$~\cite{PhysRevLett.128.040503,doi:10.1126/sciadv.adk7616,LOEM2025_supplementary}.} This transformation is implemented by setting the MHWP angles to $N\theta$ and $N\phi$, as shown in inset (b) of Fig.~\ref{figure2}.
The experimental results for $\theta = 8.5^\circ$, $\phi = 8.5^\circ$, and $N$ ranging from 1 to 10 are presented in Fig.~\ref{result2}. The plot shows $M \times \text{MSE}$ as a function of $N$ on a log-log scale, with the shot noise limits for $\theta$ and $\phi$ included for comparison. The close agreement between theoretical predictions and experimental data confirms that the LOEM strategy achieves Heisenberg scaling for multiparameter estimation.

\textit{Discussion and conclusion---}In this work, we experimentally demonstrate that the local operation with entangling measurements (LOEM) strategy saturates the QCRB in qubit systems. Furthermore, we confirm that the QFI of two parameterized mutually orthogonal states is twice that of the single-copy case [see Eq.~(\ref{QFIm1})]. For higher-dimensional systems, the QFI of parameterized mutually orthogonal states $U(\boldsymbol{x})|\psi_i\rangle$ does not necessarily equal that of $U(\boldsymbol{x})|\psi_0\rangle$ for $i = 1, \dots, d-1$. Consequently, the QFI of $|\Psi_{\boldsymbol{x}}\rangle$, the tensor product of $U(\boldsymbol{x})|\psi_i\rangle$ for $i = 0, \dots, d-1$, is not necessarily equal to $d$ times the QFI for $U(\boldsymbol{x})|\psi_0\rangle$. However, if we consider $\boldsymbol{x}$ to be completely random, i.e., with a uniform prior distribution, the average QFI over $\boldsymbol{x}$ can be used to evaluate the performance of this scheme. In this case, the average QFI is identical for all $U(\boldsymbol{x})|\psi_i\rangle$. Consequently, the average QFI of $|\Psi_{\boldsymbol{x}}\rangle$ equals $d$ times the average QFI of $U(\boldsymbol{x})|\psi_0\rangle$. This demonstrates that the LOEM strategy offers an advantage over using $d$ identical copies of the parameterized state $U(\boldsymbol{x})^{\otimes d} |\psi_0\rangle^{\otimes d}$~\cite{LOEM2025_supplementary}.

In summary, we have introduced and experimentally demonstrated the superiority of the LOEM strategy for multiparameter quantum metrology. By encoding parameters into mutually orthogonal pure states and performing entangling measurements, our approach overcomes the challenge of incompatibility and achieves the QCRB in multiparameter estimation via classical correlations, {offering a complementary approach to entangled probe state strategies that may achieve optimal precision for given unitary transformations~\cite{PhysRevLett.127.110501,PhysRevLett.115.110401}.} Using a photonic platform, we experimentally verified the advantages of the LOEM strategy, demonstrating its ability to simultaneously achieve the fundamental precision limits for the polar angle $\theta$ and the azimuthal angle $\phi$ in qubit systems. Moreover, we confirm that the LOEM strategy approaches the precision of Heisenberg scaling through iterative interactions. {Furthermore, the process for achieving $N$-fold amplification of the parameters can also be implemented using the control-enhanced sequential scheme~\cite{LOEM2025_supplementary,PhysRevLett.117.160801,PhysRevLett.123.040501,doi:10.1126/sciadv.abd2986},} further expanding the potential of quantum-enhanced measurement technologies.

\textit{Acknowledgments---}This work was supported by the Innovation Program for Quantum Science and Technology (Grant No. 2024ZD0300900), the National Natural Science Foundation of China (Grant Nos. 12347104, U24A2017, 12461160276, and 12504418), the National Key Research and Development Program of China (Grant No. 2023YFC2205802), and the Natural Science Foundation of Jiangsu Province (Grant Nos. BK20243060 and BK20233001).

M. M. and B. W. contributed equally to this work.

\textit{Data availability--}The data that support the findings of this article are openly available~\cite{Mi2025Dataset}.

\end{document}